\documentclass[11pt,twoside]{article}
\usepackage{asp2006}
\usepackage{epsf}
\usepackage{psfig}
\usepackage{lscape}
\usepackage{graphicx}

\begin{document}
\title{Protoplanetary disk fragmentation with varying radiative physics, initial conditions and 
numerical techniques}
\author{Lucio Mayer}
\affil{Institute for Theoretical Physics, University of Zurich and
Institute f\"ur Astronomie, ETH Zurich, Switzerland}
\author{Artur J. Gawryszczak}
\affil{Max Planck Institute f\"ur Astronomie, Heidelberg, Germany and
Nicolaus Copernicus Astronomical Center, Warsaw, Poland}

\begin{abstract}

We review recent results of SPH simulations of gravitational instability in gaseous protoplanetary disks,
emphasizing the role of thermodynamics in both isolated and binary systems. 
Contradictory results appeared in the literature regarding disk fragmentation at tens of AU from
the central star are likely due to the different treatment of radiation physics as well as reflecting
different initial conditions. Further progress on the subject requires extensive comparisons between 
different codes with the requirement that the same initial conditions are adopted.
It is discussed how the local conditions of the disks
undergoing fragmentation at $R < 25$ AU in recent SPH simulations are in rough agreement with
the prediction of analytical models, with small differences being likely related to the inability of 
analytical models to account for the dynamics and thermodynamics of three-dimensional spiral shocks.
We report that radically different adaptive hydrodynamical codes, SPH and adaptive mesh refinement (AMR), yield very 
similar results on disk fragmentation at comparable resolution in the simple case of an isothermal 
equation of state. 
A high number of refinements in AMR 
codes is necessary but not sufficient to correctly follow fragmentation, rather an initial resolution of the grid 
high enough to capture the wavelength of the strongest spiral modes when they are still barely nonlinear is essential.
These tests represent a useful benchmark and a starting point for a forthcoming code comparison with realistic radiation physics.  

\end{abstract}

\section{Overview}

Physical fragmentation in astrophysical systems such as gravitationally unstable protoplanetary disks depends on the competition between gravity and thermal pressure, at least in the limit in which the contribution of magnetic fields is neglected. 
Knowing whether existing numerical simulations are modeling correctly both gravity and pressure is then of paramount 
importance in this context. Thermal pressure is affected by the details of cooling and heating, either by 
radiation, convection or viscosity. The increasingly more sophisticated simulations designed in the last few years have explored the effect of  thermodynamics on disk fragmentation.
finding in general that this is less likely than in older models in which the gas was evolved using a
locally isothermal equation of state (Boss 2002a,b; Mayer et al. 2002; 
Rice et al. 2003). Currently, it is debated whether fragmentation into Jupiter-sized clumps
is possible, especially at distances less than $50$ AU from the central star (Durisen et al. 2007; Stamatellos \& Whitworth 2007).
It is well understood that the disk must cool on a timescale
comparable to the orbital time in order for fragmentation to occur (Rice et al. 2003, Gammie 2001; Johnson \& Gammi 2003;
Clarke et al. 2007). At a few tens of AU 
the optically thick disk midplane  cools via radiation on a timescale longer than the local 
orbital time, hence
the only chance for disks to cool efficiently is via a non-radiative mechanism. This 
could be either convection
or turbulent diffusion associated with shock bores (Boss 2004; Mayer et al. 2007; Boley \& Durisen 2006). Finally, analytic calculations that include 
convection predict no fragmentation 
at radii $< 50$ AU for disks with masses significantly smaller than the mass of the central star, as it is the case in T Tauri disks (Rafikov 2005, 2007).

Four groups  have implemented
a scheme for radiative transfer in three dimensional simulations. Two of them, using, respectively,
an SPH and a finite-difference polar grid code, and similar initial conditions, find that fragmentation can 
happen at  $R < 20$ AU (Boss 2004; Mayer et al. 2006),
while the  two other groups use , respectively, an SPH (Stamatellos \& Whitworth 2007)
and a cylindrical grid code (Boley et al. 2006) with nearly
identical initial conditions and find that disk fragmentation does not happen at $R < 50$ AU
(Stamatellos \& Whitworth 2007 find that fragmentation is possible at larger radii, $R \sim 100$ AU).
These contradictory results were obtained with four different codes. The two sets of initial conditions
were also significantly different, the disks in Mayer et al. (2007) and Boss (2006) having much higher
surface densities than those in Boley et al. (2006) and Stramatellos \& Whitworth (2007), and thus being
more prone to fragmentation.  Numerical codes differ not only in the way they implement radiative
transfer but also in more basic aspects of the algorithm such as how they compute gravity and how they solve the energy
equation. Further progress in elucidating whether disk instability is a  possible formation mechanism
for giant planets clearly require that the different codes adopted in this area are compared on
identical conditions, first in simple models with a fixed equation of state and then 
on progressively more sophisticated models with radiative transfer. 
Here we summarize the results recently obtained with the SPH code GASOLINE (Wadsley, Stadel \& Quinn 2004)
with a scheme for radiative transfer for 
the case of both isolated and binary systems. Then we discuss the results of the first stage of a code comparison
project in which the isothermal disks are evolved with GASOLINE and with one of the most advanced grid codes currently
available, the adaptive mesh refinement (AMR) code FLASH (Fryxell et al. 2000).

\section{Flux-limited diffusion simulations with GASOLINE}

In Mayer et al. (2007) we have performed SPH simulations using a flux-limited diffusion solver recently
implemented in GASOLINE. The scheme follows from Cleary \& Monaghan (1997), uses the flux-limiter by Bodenheimer
et al. (1990) and includes tabulated Rosseland mean opacities by d'Alessio et al. (1997). 
The edge of the disk cools as a blackbody.
To find particles that are ``on the edge'' of the disk, we examine the directions to all of the neighbors used in 
smoothing sums. If a particle has no neighbors within a certain fraction of a solid angle 
from a preferred direction (edge detection angle, EDA) it is considered an edge particle.
From the geometry of a disk, the preferred directions (treated independently) are out of the plane of the disk 
(both up and down) and radially outward.

Hence to model the radiation from edge particles we add the following term to the energy equation of each 
particle
\begin{equation}
\label{eqn:udot_black_body}
\dot{U}_a = f_a S \sigma T_a^4 / m_a .
\end{equation}

{\noindent where $S=4\pi h_a^2$ is the surface through which the 
particle radiates, with $h_a$ the smoothing length of the particle.
The ``edgeness factor'' $f_a$ represents the fraction of their surface area over which a particle radiates. It is 
usually zero, and takes value $1/2$ 
for particles on one of the up, down, or out boundaries, $1$ for those
on the edge in both the up, down and out directions, and $3/4$ for those on the 
edge in the out and either up or down direction. More details on the
scheme can be found in Mayer et al. (2007).

Several simulations with different values of EDT are performed, with  $\mu = 2.4$, an adiabatic index $\gamma=1.4$, and 
an opacity consistent with solar metallicity.
Some simulations are restarted with a higher molecular weight after spiral shocks begin to
develop. In the spiral shocks a molecular weight higher than solar could result from the
combination of three factors, i.e. a higher mean metallicity of the gas, an increased dust-to-gas ratio
and the vaporization
of dust grains. Water ice can be vaporized in the spiral shocks of a gravitationally
unstable disk, where temperatures can rise above $150$ K (Nelson et al. 2000; Mayer et al. 2005)
and accounts for about $30-40\%$  of the dust content. The dust/gas ratio is expected to
increase by an order of magnitude in the spiral arms as 
grains larger than micron size rapidly migrate towards gaseous overdensities (Rice et al. 2004; 
Haghighipour \& Boss 2003). 
An order of magnitude enhancement in the gas-to-dust ratio and the vaporization of water ice would 
produce $\mu \sim 2.5$ in a disk with mean solar metallicity, while $\mu \sim 2.85$ would be achieved in 
a disk  with a mean metallicity three times  larger than solar.

\begin{center}
\begin{figure}[!h]
\plotone{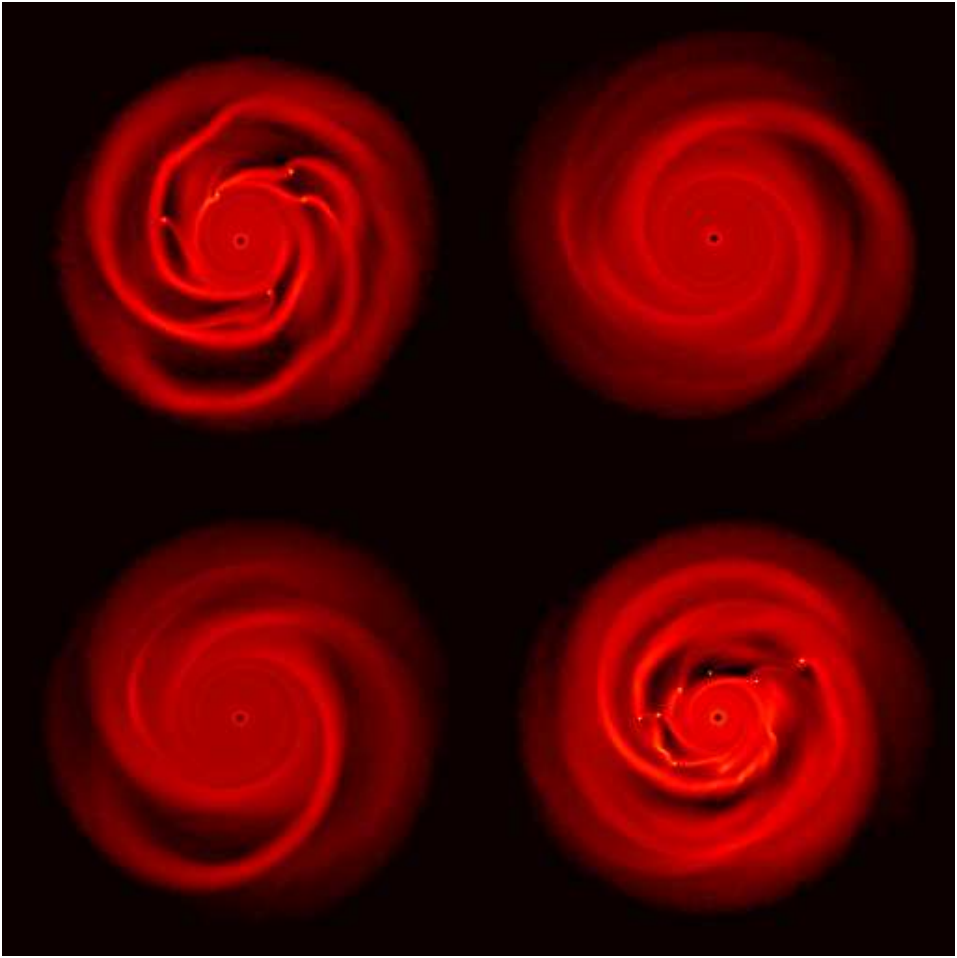}
\vspace{0in} 
\caption{{ Color coded logarithmic density plots of the disk seen face-on after 1300
years of evolution. We show the runs with $\mu=2.4$, EDT (edge detection angle)
=30 degrees (top left), $\mu=3$, EDT=60 degrees (top right), $\mu=2.4$, EDT=40 degrees (bottom left)
and $\mu=2.7$, EDT=40 degrees (bottom right). The maximum densities shown, at the clumps position, reach
values higher than $10^{-9}$ g/cm$^3$.}}
\end{figure}
\end{center}

The disk grows uniformly in mass at the rate of $\sim 10^{-4} M_{\odot}/$yr, approaching $0.1 M_{\odot}$ 
after about $10^3$ years. 
When its mass grows above $0.05 M_{\odot}$ the Toomre parameter $Q$ drops below 2 in the outer part of
the disk and strong spiral patterns begin to appear. 
The shocks occurring  along the spiral arms limit the growth of their amplitude as the increasing
pressure counteracts self-gravity. Yet, fragmentation occurs at $R \sim 15$ AU
in some of the simulations once the disk mass is in the range $0.12-0.15 M_{\odot}$ (Figure 1).

Previous works have shown that fragmentation happens only if the disk is able to cool on a timescale comparable
to the local orbital time (up to 50\% longer for runs in which $\gamma=1.4$ according to Mayer et al. (2005) 
and Lodato \& Rice (2005)).
The disk midplane heats to temperatures larger than $200$ K as a result of spiral shocks, these being
also the sites more favourable to fragmentation based on the increased local density. In runs in
which fragmentation happens we observe vertical gas motions, due to either convection of shock bores
(Boley et al. 2006), fast
enough to redistribute thermal energy on the the orbital timescale at $10-15$ AU, 
$30-50$ years. Such cooling time is short enough for fragmentation
to happen (Mayer et al. 2004a; Rice et al. 2003) provided that the radiation can
leave the disk over a timescale also shorter than the orbital time.
It is then not surprising that fragmentation depends on the value of EDT in our runs
which determines the ultimate cooling rate of the disk once the heat generated by compressional
and viscous heating has been transported to the disk surface (see Figure 1).
We also evolved some disks with an opacity reduced by a factor of 10-50 and found no difference, 
in agreement with the findings of Boss (2002b). This is because that radiative cooling rate
in the optically thick part of the disk, which is controlled by opacity, remains smaller than
the compressional heating rate even for small opacities, which is further evidence that heat ought
to be transported from the midplane to the surface by some other mechanism.

\subsection{Dependence on atmospheric cooling}

Disk fragmentation is seen to depend sensibly on how fast the disk atmosphere cools, which highlights
the importance of modeling correctly the net radiative loss at the disk boundary (Boley et al. 2006).
Disks fragment more easily for smaller EDTs, which effectively correspond to a larger emitting surface area
and hence to a larger cooling rate. Small changes in EDT
(e.g. from 60 to 40 degrees, corresponding to an increase in the emitting surface area of less than 50\%) 
can bring the disk from fragmentation to thermal self-regulation and subsequent stability.
At a fixed height above the midplane different regions of the disks can have temperatures different by
up to a factor of 4, with hotter
regions along spiral shocks often surrounded by much colder regions. This is not surprising since the 
disk has low optical depths over a larger fraction of its
vertical height in such underdense inter-arm regions.
The different optical depths profiles in these two types of regions can be seen in Figure 2. The study of the vertical structure and local variations in optical depth warrants further investigation and strongly suggests that the cooling is really local in these disks. On the other end it appears that the colder gas surrounding the spiral shocks accretes onto the midplane 
overdensities in the arm, contributing to their growth. This shows that just the knowledge of the local thermodynamical conditions, such as the local cooling
time, is not enough to understand the nonlinear evolution of overdensities. The latter is an example of
an aspect of disk instability that only three-dimensional simulations can describe properly.

\subsection{Dependence on molecular weight}

We find that $\mu \ge 2.4$ is a  necessary condition for fragmentation (Figure 1). Larger molecular weights
can have two effects; they lower the value of the pressure gradients in the adiabatic compression
term of the internal energy equation, since for an ideal gas $P \sim T/\mu$ when $\mu$ is increased
while holding $T$ fixed,  and they increase the cooling rate at the surface (see eq. 1) since 
$T \sim \mu$ when $P$ is held fixed. 
We verified that it is the first effect that dominates the effect of changing the molecular weight (see Mayer et al. 2007).
The fact that even a small reduction of compressional heating can change the result is not suprising; the same
effect is seen when the adiabatic index is change between $5/3$ and $7/5$, as was previously shown (Rice, Lodato
\& Armitage 2005; Mayer et al. 2005). The fact that the system is so sensitive to the details of heating
and cooling shows once more why it is crucial to compare carefully the different implementations
in the different codes. It also suggests that a self-consistent description of
how the molecular weight and the adiabatic index should change in space and time is desirable in future
simulations. The implementation of a variable adiabatic index that reflects the different states of molecular
hydrogen at different temperatures (Boley et al. 2007) is a first important step in this direction.

 \begin{figure}[!h]
\plotone{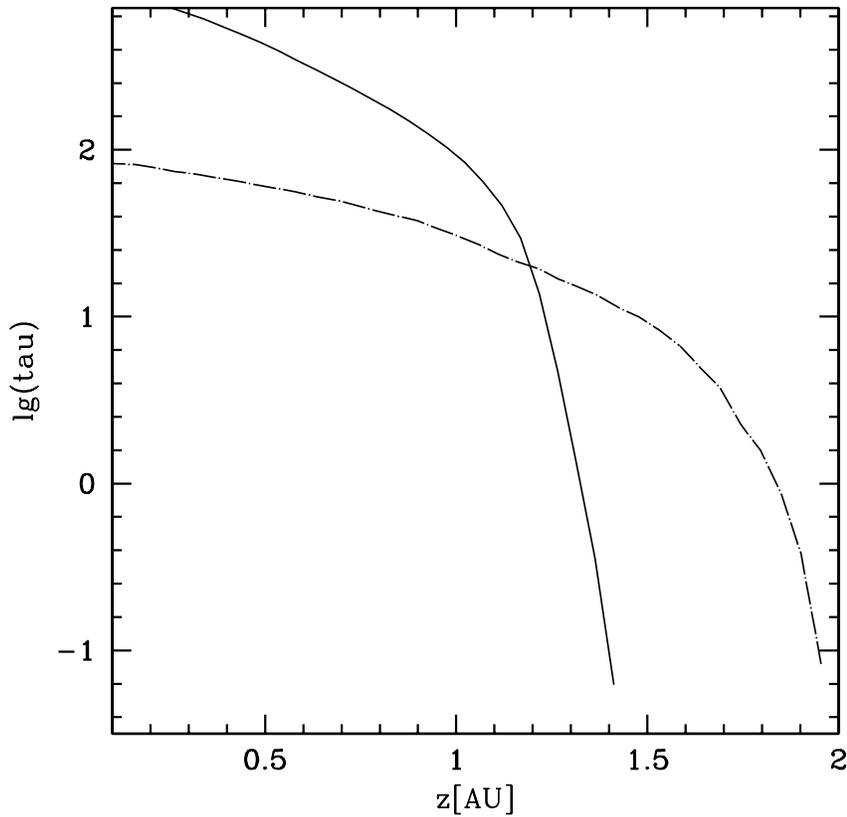}
 \caption{{ Optical depth profiles as a function of the height above the disk midplane in a region within 
a spiral shock that will later produce a clump (solid line) and in a neighboring, underdense region between spiral 
arms (dot-dashed line). The optical depth is azimuthally averaged in a cylindrical column that cuts through a 
selected region of the disk, the main axis of the cylinder being perpendicular to the midplane.}}
 \end{figure}

\subsection{Comparison with analytical predictions and other comments}

We compared the conditions of the disks in the simulations that lead to fragmentation in Mayer et al. (2007)
with the analytical predictions of Rafikov (2007), which include convective cooling. We find that
the typical midplane surface densities of our disks in the spiral arms just before fragmentation, $2500-3000$ 
g/cm$^3$, are sefely above the minimum surface densities necessary for fragmentation predicted by
Rafikov at distances of about $20$ AU, and the same applies to the optical depth at the midplane, which is 
$\tau \sim 10^3$ in our simulations. At the same radii midplane temperatures in spiral shocks 
before fragmentation reach $300$ K, about a factor of 2 lower than those predicted by the analytical model assuming $Q_0 = 1.4$ for the threshold of fragmentation and $\mu =2.7$ (simulations with $\mu =2.7$ require less restrictive
choice of EDT in order to fragment). While such a mismatch
between midplane temperatures has to be investigated further, if we take into account that the
analytical models do no treat a crucial aspect of disk thermodynamics, namely shock
dissipation in three dimensions, the overall agreement is quite satisfactory. We emphasize that in the
simulated disks the disk surface density profile evolves away from the simple power-law as a result of
transport of angular momentum by spiral modes (Mayer et al. 2005), explaining why very high 
local surface 
densities can coexist with a total disk mass $< 0.2 M_{\odot}$, almost an order of magnitude lower than
that predicted by analytic models assuming the standard density profile of the minimum solar nebula 
(Rafikov 2005). The masses of the simulated disks are thus still compatible with the most massive
among T Tauri disks (Beckwith et al. 1990). Nevertheless, these massive disks, and thus 
gravitational instability, are likely more
representative of the earlier stages of disk evolution (Class 0 to Class 1), currently almost 
unexplored by observations. 

Recently Nelson (2006) has emphasized the importance of resolving the vertical structure of the disk 
in fragmentation simulations. Our edge detection scheme is in principle resolution dependent 
since the surface of identified edge particles, that appear explicitly in the cooling calculation, depend on the
local SPH smoothing length. Recently we have run again one simulation with up to 6 times more particles, 
for a total of $6 \times 10^6$ particles, now the highest resolution SPH disk fragmentation simulation
available, and
found that fragmentation is even more vigorous compared to the original run (with EDA = 40 degrees and 
$\mu = 2.7$).

\section{Disk fragmentation in binaries}

When a massive disk ($0.1 M_{\odot}$) around a solar mass star interacts with a companion of comparable mass at separations of less than $60$ AU a cooling time 2 to 3 times shorter than the orbital time 
is needed 
to counteract shock heating along the spirals arms and allow fragmentation 
(see Mayer et al. 2005). Tidally forced spiral arms reach indeed a much higher amplitude relative 
to their counterparts in isolated disks, and hence shock heating in the arms is stronger.
Such cooling times are too short to be realistic, hence Mayer et al. (2005) concluded that fragmentation
is suppressed in tightly bound binaries, a separation $> 100$ AU being necessary for the disk to
behave as in the non-interacting case. Instead, in disks with intermediate masses ($0.05$ $M_{\odot}$)
the lower self-gravity was found to compensate for the tidal forcing,
producing less shock heating and allowing fragmentation for more realistic cooling times, comparable 
with the orbital time. However, newer calculations with flux-limited
diffusion show that the same disks cannnot cool fast enough and thus do not fragment
(Mayer, Boss \& Nelson 2007). Conversely,
Boss (2006) finds that disk in binaries with separation of about $50$ AU fragment more easily than in isolation, 
even when radiative transfer is included. The comparison between these different works is complicated by the fact
that different types of initial conditions are used, namely disks that are not fragmenting
in isolation are chosen by Boss (2006) for the most part while Mayer et al. (2005) studied predominantly disks that
were known to fragment in isolation. Perhaps heating by artificial viscosity present in the SPH
calculations of Mayer et al. (2005) plays a role in suppressing fragmentation, although the 
good match between SPH and AMR calculations (see next section) 
suggests that artificial viscosity can hardly have a major effect in high resolution calculations.
We defer the reader to the recent review by Mayer, Boss \& Nelson 2007 for a detailed discussion of fragmentation
in binaries.

\vskip1.5truecm

 \begin{figure}[!h]
\plotone{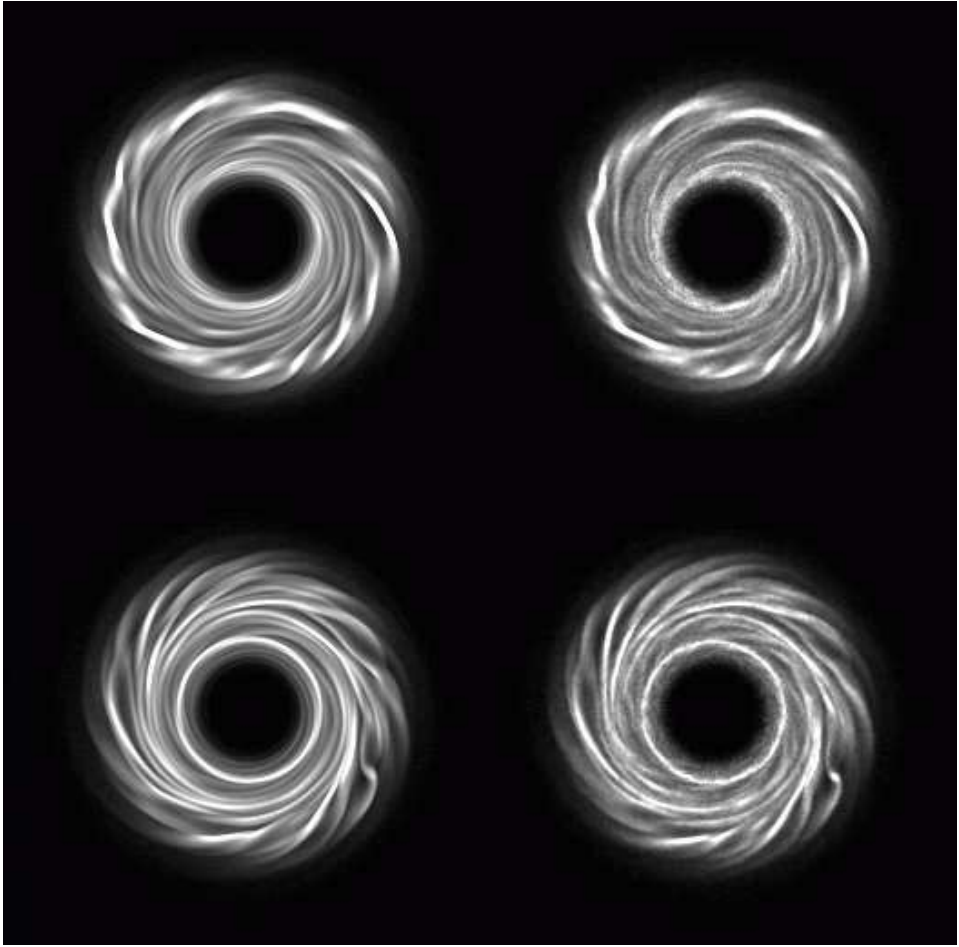}
 \caption{{ Grey-scale logarithmic density plots showing the disk in a run
with FLASH (left) and a run with GASOLINE (right) with comparable spatial
resolution ($n_{ref}=8$ was used in FLASH, see text). The top panel show
the disks after 1.5 orbital times (orbital time measured at 20 AU) and the bottom
panel shows the disks at 2.5 orbital times, when a clump has formed in
both cases at nearly identical positions (at about 4 o'clock.)}}
\end{figure}

\section{The Wengen code comparison: GASOLINE vs. FLASH}.

In this section we report on how GASOLINE and FLASH were used to evolve identical disk initial conditions with 
varying resolution and a locally isothermal equation of state. GASOLINE (Wadsley, Stadel \& Quinn 2004)
is a parallel hydrodynamics code that uses a fairly standard implementation of SPH with the Monaghan form
for artificial viscosity and the Balsara switch to reduce viscosity in shear flows. It performs smoothed
sums over a fixed number of neighbors (=32), solves the
energy equation using the asymmetric form and gravity is computed using a binary tree. It adopts a fixed
particle gravitational softening and a leapfrog integrator with multiple hierarchical timesteps.
The FLASH code (Fryxell et al. 2000) is a finite difference code which employs parallelized,
block-oriented Adaptive Mesh Refinement (AMR) technique. The hydrodynamical equations are solved using PPM 
coupled with an isothermal Riemann solver in the tests shown here.
Refinement was done whenever the number of cells per Jeans length, $n_{\mathrm{J}}$ has dropped below a fixed threshold $n_{\mathrm{ref}}$.
For practical reasons the maximum level was limited, so 
regions where density was higher than

The refinement criterion was such that $n_{\mathrm{ref}} = 8$, and the resolution was 
limited to $\min \Delta x = \min \Delta y = 25/1024$ AU, $\Delta z = \Delta x/2$. The computational domain was
a periodic box $[-50\,\mathrm{AU}, 50\,\mathrm{AU}]^3$. The choice of $n_{\mathrm{ref}} = 8$
guaranteed that resolution in the volume occupied by the disc at $t = 0$, was $\Delta x = 25/256$ AU, comparable
to the initial spatial resolution of the SPH calculation (a lower resolution initial grid was used in one of the
runs to study and the results are shown in Fig. 4).
Note that this is twice as much as required according to Truelove et al. (1997).

\vskip 1.5truecm

\begin{center}
\begin{figure}[!h]
\plotone{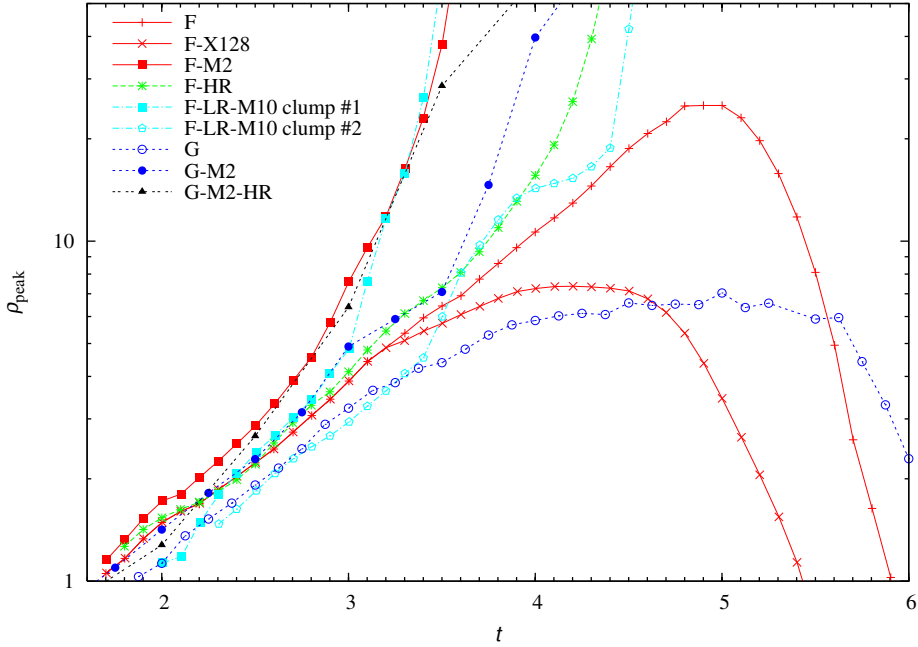}
 \caption{{Maximum density of the first clump that forms as a function of time in a subset of the
runs performed. Time is in simulation units (1 disk orbital time at 10 AU corresponds to $\Delta t \simeq 4$). 
Letter F and G stand for FLASH and GASOLINE runs, respectively. Strings 'M2' 
and 'M10' in run names indicate that initial density was increased uniformly 
by 2\% and 10\%, respectively, 'X128' means that maximum resolution was limited 
to $\Delta x = 25/512$ AU, 'HR' and 'LR' for FLASH runs define refinement 
criterion $n_{\mathrm{ref}} = 12$ and 4, respectively, 'HR' for GASOLINE runs 
indicates that number of particles was $1.3 \times 10^6$ instead of $1.65 \cdot10^5$. 
Density $\rho_\mathrm{peak} = 1$ corresponds to $2.4\cdot10^{-10}$ g/cm${^3}$.
}}
  \label{fig:FGdenspeak}
 \end{figure}
\end{center}

The base simulations evolved a 3D isothermal disk with an initial $Q_\mathrm{min} \sim 1.0$ (i.e.  larger than the critical
$Q_\mathrm{min} = 0.67$ for fragmentation in three dimensions, see Nelson et al. 2000). The disk has a mass of $0.055\,M_{\odot}$ and a temperature of 20 K, and extends from 4 to 25 AU.
A similar comparison for a single
resolution has been shown in Durisen et al. (2007). There the disk had a slightly lower initial $Q_{min} \sim
0.8$ and was fragmenting after just one orbital time for all codes. The fact that the disk starts somewhat below 
the threshold for fragmentation in the new runs discussed here made it harder to reach a reasonable convergence
between the two codes  since the initial marginally unstable state can be rapidly stabilized or destabilized 
depending on the details of how the  density evolves in the first one or two orbital times. In other words, small
perturbations of the state produced by small differences in the forces at play, gravity, pressure and, in the
case of SPH, artificial viscosity, can produce rapidly diverging evolutionary trajectories when different
implementations are used. It is then remarkable that, when the spatial resolution is sufficiently high to resolve
the wavelength of the most unstable modes in both codes, the results given by AMR and SPH are
very similar (Fig. 3).
The power in the various modes obtained via Fourier decomposition of the density field was
also found to be almost identical.
The degree of similarity is such that the first clump has nearly the same location and characteristic density 
in GASOLINE and FLASH runs with comparable spatial resolution both at the beginning
and during the simulation (Fig. 3).
The peak densities also differ by less than $30\%$ at comparable resolution
(Fig. \ref{fig:FGdenspeak}, compare ``F'' and ``G'' runs).  
Clumps are indeed slightly sharper (i.e. have higher peak densities) in the FLASH runs,
perhaps due to the lower artificial viscosity. This suggests that, contrary to previous claims but in agreement with
the analysis of Mayer et al. (2004), artificial viscosity suppresses rather than enhances fragmentation.

For both codes results also change in a similar way with increasing resolution since the density profile
of the clumps becomes progressively steeper (this is shown by the increasing peak density reported in
Fig. \ref{fig:FGdenspeak}). The latter result is not surprising given the fact that an
isothermal equation of state is adopted -- it is in some sense as an artifact of such idealized
thermodynamical assumption. 
In reality the clumps will become nearly adiabatic at sufficiently high density and 
their density profile should converge with increasing resolution as such density is approached.

After several orbital times small differences in the computation of the density field produce 
an increasingly different pattern of fragmentation and convergence would have to be sought in a statistical
sense, i.e. comparing a large set of initial conditions rather than a single specific case.
The agreement between the two codes persist when we perturb slightly the initial state of the disk by
increasing or decreasing the initial mass by a few percent (Fig. \ref{fig:FGdenspeak}), suggesting that substantial
convergence has been reached at least for the first few orbital times of the disk evolution.

Another important result that we obtained is that the {\it initial} spatial resolution, namely the grid spacing in AMR
or the initial gravitational softening in SPH, affects significantly the outcome.
For example, in the FLASH runs many refinements are
useless if the initial grid is coarse and is not able to resolve the wavelength of the strongest modes well enough
from the beginning (Fig. 4, compare runs ``F'' and ``F-X128').
Resolving the single most unstable mode only might not be enough if the nonlinear evolution
is partially determined by mode coupling (see Laughlin, Korchagin \& Adams 1998).
A similar effect is seen in SPH if the gravitational softening is initially too large, independently on whether it is well
matched or not with the smoothing length. This implies that SPH simulations adopting adaptive softening are
not necessarily better than simulations with fixed softening because they also need a high enough mass resolution
to enforce that the softening be smaller than the wavelength of the relevant disk modes since
the beginning of the calculation.
This requirement is complimentary to the Bate \& Burkert (1997) resolution criterion based on resolving 
the local Jeans length, and will in general be much stricter since the wavelength of the strongest non-axisymmetric
modes is typically shorter than the local Jeans length. We will attempt to formalize the new criterion in a forthcoming paper.

The excellent match betwen FLASH and GASOLINE suggests that in circumstances in which self-gravity plays a 
crucial role in the dynamics and thermodynamics of a system, SPH is
a robust technique which is not more diffusive than a PPM code despite the use of artificial viscosity, at
least at the fairly high resolutions adopted here (from $10^5$ to $> 10^6$ particles). It is noteworthy that GASOLINE
runs were a factor of 10 less time consuming than FLASH runs at comparable resolution, mostly thanks to the
faster gravity calculation. In other types of
astrophysical situations, such as those involving hydrodynamical instabilities, AMR codes capture the
physics more correctly compared to standard SPH (Agertz et al. 2007).

\section{Final remarks}

The two main obstacles towards establishing clearly how likely is that giant planets form by gravitational instability are
the fact that all the radiation physics models introduced so far rely on a number of simplifications and the fact that 
the initial
conditions of protoplanetary disks are poorly constrained. The two issues partially overlap. Indeed disks are evolved as 
if they were in isolation while both their formation and their evolution will be affected by the properties of the collapsing
molecular cloud core, including the local radiation environment and magnetic fields. 
External irradiation can heat the disk and stabilize it (Boley et al. 2006), but its overall effecton disk thermodynamics could be even more
subtle. For example, if the molecular envelope has been mostly dispersed strong UV and X-ray heating from nearby young massive stars could penetrate down to the disk, increasing its temperature 
towards the opacity gap at $T \sim 10^3$ K, thus triggering rapid cooling and possibly fragmentation as in the scenario envisioned by Johnson \& Gammie (2003).

A lumpy collapse, as expected if cores are turbulent, could also favour disk instability if massive gas clumps
hit the disk episodically. In addition, cores themselves are not isolated systems. In some
models of star formation dynamical interactions between cores can be frequent and can translate into interactions between the disks themselves. Therefore there is a need of multi-scale simulations that study the formation and evolution of disks starting from the collapse of turbulent cores. Magnetic fields also will play a role in setting the initial conditions, and their effect on disk evolution, in particular whether they favour or disfavour the onset of gravitational instability, is hard to predict without simulations that incorporate simultaneously MHD, self-gravity and realistic radiation physics (see Fromang 2005 on first attempts to study the combined effect of magnetorotational 
and gravitational instabilities in isothermal protoplanetarty disks).
Considering that all this is still missing in the models it is quite possible that we have just begun to understand the role of gravitational
instability in planet formation both as a possible mechanism to form giant planets and as a key process
responsible for the evolution of protostellar disks. 
These are issues relevant to any model of planet formation.

\section*{Acknowledgments}
The AMR software (FLASH) used in this work was in part developed by the
DOE--supported ASC / Alliance Center for Astrophysical Thermonuclear
Flashes at the University of Chicago. 
Simulations were performed on the Zbox2 cluster at the University of Zurich,
on the Gonzales cluster at ETH Zurich and on the PIA cluster at the Rechenzentrum Garching.
LM was supported by a grant of the Swiss National Science Foundation.
AJG was supported by Alexander von Humboldt Foundation and
by Polish Ministry of Science through grant 1 P03D 026 26.

\end{document}